\documentclass[prd]{revtex4}

\usepackage{amsmath}
\usepackage{graphicx}

\begin{document}

\title{Non-contact gears: II. Casimir torque between 
concentric corrugated cylinders for the scalar case}
\author{In\'{e}s Cavero-Pel\'{a}ez}
\email{Ines.Cavero-Pelaez@spectro.jussieu.fr}
\affiliation{Laboratoire Kastler Brossel,
Universit\'{e} Pierre et Marie Curie, ENS, CNRS, 
Campus Jussieu,
University Paris 6, Case 74, F-75252 Paris, cedex 05, France.}
\author{Kimball A. Milton}
\email{milton@nhn.ou.edu}
\homepage{http://www.nhn.ou.edu/
\author{Prachi Parashar}
\email{prachi@nhn.ou.edu}
\author{K. V. Shajesh}
\email{shajesh@nhn.ou.edu}
\homepage{http://www.nhn.ou.edu/
\affiliation{Oklahoma Center for High Energy Physics
and Homer L. Dodge Department of Physics and Astronomy,
University of Oklahoma, Norman, OK 73019, USA.}

\date{\today}

\begin{abstract}
The Casimir interaction between two concentric corrugated cylinders 
provides the mechanism for non-contact gears. To this end, we 
calculate the Casimir torque between two such cylinders, described by 
$\delta$-potentials, which interact through a scalar field. We derive 
analytic expressions for the Casimir torque for the case when the 
corrugation amplitudes are small in comparison to the corrugation 
wavelengths. We derive explicit results for the Dirichlet case, and 
exact results for the weak coupling limit, in the leading order. The 
results for the corrugated cylinders approach the corresponding 
expressions for the case of corrugated parallel plates in the limit 
of large radii of cylinders (relative to the difference in their 
radii) while keeping the corrugation wavelength fixed.
\end{abstract}

\maketitle

\section{Introduction}

The Casimir torque between two material bodies, which is the 
rotational analog of the Casimir force~\cite{Casimir:1948dh}, 
was studied for the first time in 1973~\cite{Barash:1973}. 
The Casimir torque between two uni-axial birefringent dielectric plates
has been studied more recently in \cite{Enk:1995}.
The lateral Casimir force between corrugated parallel plates was 
calculated perturbatively in \cite{Emig:2001dx}.

Lately a non-contact rack and pinion arrangement has been
proposed in \cite{Ashourvan:2007} and discussed in the 
proximity force approximation (PFA) limit.
This proposal has been generalized to the design of a non-contact gear
consisting of two corrugated concentric cylinders in 
\cite{lombardo-2008} to discuss possible experimental arrangements. 
The perturbative calculation for the Casimir torque between 
two concentric corrugated cylinders, in the spirit of~\cite{Emig:2001dx},
has not yet been carried out. We achieve this task to the leading 
order for the scalar case here.  The next-to-leading-order calculation 
for the corrugated cylinders, in the spirit of the corresponding 
case of corrugated plates~\cite{Cavero:}, is in progress. 
 
\subsection{Statement of the problem}

We consider two concentric, semi-transparent,
corrugated cylinders described by the potentials,
\begin{equation}
V_i(r,\theta) = \lambda_i \,\delta (r - a_i - h_i(\theta)),
\end{equation}
where $i=1,2$ are labels that identify the individual cylinders,
and we shall have $a = a_2 - a_1 > 0$.
The functions $h_i(\theta)$ describe the corrugations associated
with the cylinders.
We define the function 
\begin{equation}
a(\theta) = a + h_2(\theta) - h_1(\theta),
\label{a-th}
\end{equation}
which measures the relative corrugations between the cylinders.
We shall define the corrugations $h_i(\theta)$ such that the 
mean of the relative corrugations evaluates to $a$,
\begin{equation}
\frac{1}{2\pi} \int_0^{2\pi}
d\theta \,\big[ h_2(\theta) - h_1 (\theta) \big] = 0.
\end{equation}

Let $E$ be the total Casimir energy associated with the 
two concentric corrugated cylinders, including the divergent contributions
associated with the single cylinders. In general this energy changes
if we rotate one of the cylinders with respect to the other and thus
leads to a torque between the cylinders.
This shift can be described by an angular rotation, $\theta_0$,
of the corrugations on the inner cylinder as
$h_1(\theta + \theta_0)$ and the corresponding torque will be
\begin{equation}
{\cal T} = - \frac{\partial E}{\partial \theta_0}.
\label{torE}
\end{equation}

We observe that due to rotational symmetry 
there will be no torque between two uncorrugated cylinders.
Further, using the same argument there will be no torque 
when only one of the cylinders has corrugations on it.
In light of these observations it is helpful to write the total 
Casimir energy for two corrugated concentric cylinders
in the form
\begin{equation}
E = E^{(0)} + E_1 + E_2 + E_{12}(\theta_0),
\label{energy}
\end{equation}
where, $E^{(0)}$ is the energy of the configuration
when the corrugations are absent on both the cylinders,
$E_i$ is the additional contribution to the Casimir energy when one of 
the cylinders is uncorrugated, and 
$E_{12}$ is the contribution to the energy which is present only
when both cylinders are corrugated.
Thus, $E_{12}$ is the interaction energy due to the presence of 
corrugation, and only this term in eq.~\eqref{energy} contributes
to the torque between the cylinders.
The physical quantities associated with the uncorrugated cylinders 
thus act as a background, and a reference, and we shall find it 
convenient to denote them by the superscript $(0)$ to mean 
zeroth order. The potential for the background is
\begin{equation}
V_i^{(0)}(r) = \lambda_i \delta (r - a_i),
\end{equation}
which has no angular dependence.

Using the multiple scattering formalism, see 
\cite{Emig:2007me,Milton:2007wz} and references in \cite{Milton:2007wz}, 
which has been further discussed for the case of corrugated plates
in \cite{Cavero:}, we can write
\begin{equation}
\Delta E = E_1 + E_2 + E_{12},
\label{DE=DEi+}
\end{equation}
where in accordance with the notation given in \cite{Cavero:},
$\Delta X$ represents the deviation of the physical quantity, $X$, 
from the background.
The term in eq.~\eqref{DE=DEi+} which contributes to the Casimir torque,
to the leading order, is given by~\cite{Cavero:}
\begin{equation}
E_{12}^{(2)} = \frac{i}{2\tau} \,\text{Tr}
\Big[ G^{(0)} \Delta V_1^{(1)} G^{(0)} \Delta V_2^{(1)} \Big],
\label{dE12}
\end{equation}
where the Green's function associated with the
background satisfies the differential equation,
\begin{equation}
\Big[ -\partial^2 + V_1^{(0)} + V_2^{(0)} \Big] G^{(0)} = 1.
\label{G0eqn}
\end{equation}
In the leading order the corrugations are treated as small perturbations,
$|h_i(\theta)| \ll a < a_i$, 
which lets us approximate the potentials by 
\begin{equation}
\Delta V_i(r,\theta) \approx V_i^{(1)}(r,\theta)
= h_i(\theta) \frac{\partial}{\partial r} V_i^{(0)}(r),
\end{equation}
where we have used the superscript $(1)$ to represent the
first order perturbation in the quantity.

\section{Green's function}

We observe that the evaluation of the interaction energy,
$E_{12}$, to the 
leading order, involves solving for the Green's function for the
configuration involving the background alone.
This amounts to solving the differential equation in
eq.~\eqref{G0eqn}, whose solution can be written as
\begin{equation}
G^{(0)}(x,x^\prime)
= \int \frac{d\omega}{2\pi} \,e^{-i\omega (t - t^\prime)}
  \int \frac{dk}{2\pi} \,e^{i k (z - z^\prime)}
  \sum_{m=-\infty}^{\infty} \frac{1}{2\pi} \,e^{im(\theta - \theta^\prime)}
  \,g_m^{(0)}(r,r^\prime;\kappa),
\end{equation}
where $\kappa^2=k^2 - \omega^2$. The reduced Green's function,
$g_m^{(0)}(r,r^\prime;\kappa)$, satisfies the equation
\begin{equation}
-\left[ \frac{1}{r} \frac{\partial}{\partial r} r \frac{\partial}{\partial r}
  - \frac{m^2}{r^2} - \kappa^2
  - \lambda_1 \delta (r-a_1) - \lambda_2 \delta (r-a_2) \right]
g_m^{(0)}(r,r^\prime;\kappa)
= \frac{\delta (r-r^\prime)}{r}.
\label{g0rrp}
\end{equation}
The solution for $g_m^{(0)}(r,r^\prime;\kappa)$
in the above equation 
is expressed in terms of the modified Bessel functions as
\begin{align}
g_m^{(0)}(r,r^\prime;\kappa) =
\begin{cases}
I_m(\kappa r_<) K_m(\kappa r_>) - \frac{1}{\Delta} 
\big[ \lambda_1 a_1 \lambda_2 a_2 \, K_1 K_2 \, \big( I_2 K_1 - I_1 K_2 \big)
\\[2mm] \hspace{45mm} + \lambda_1 a_1 \, K_1^2 + \lambda_2 a_2 \, K_2^2 \big]
\,I_m(\kappa r) I_m(\kappa r^\prime),
& \text{if} \quad (r,r^\prime) < a_1 < a_2, \\[3mm]
I_m(\kappa r_<) K_m(\kappa r_>) + \frac{1}{\Delta} 
\big[ -\lambda_2a_2 \,K_2^2 \,\big(1 +\lambda_1 a_1 \,I_1 K_1 \big)\big]
\,I_m(\kappa r) I_m(\kappa r^\prime) \\[2mm]
\hspace{25.0mm} ~~ + \frac{1}{\Delta} 
\big[ -\lambda_1a_1 \,I_1^2 \,\big(1 + \lambda_2 a_2 \,I_2 K_2 \big)\big]
\,K_m(\kappa r) K_m(\kappa r^\prime) \\[2mm]
\hspace{25.0mm} ~~ + \frac{1}{\Delta} 
\big[ \lambda_1a_1 \lambda_2a_2 \,I_1^2 K_2^2 \big]
\,I_m(\kappa r) K_m(\kappa r^\prime) \\[2mm]
\hspace{25.0mm} ~~ + \frac{1}{\Delta} 
\big[ \lambda_1a_1 \lambda_2a_2 \,I_1^2 K_2^2 \big]
\,K_m(\kappa r) I_m(\kappa r^\prime),
& \text{if} \quad a_1 < (r,r^\prime) < a_2, \\[3mm]
I_m(\kappa r_<) K_m(\kappa r_>) - \frac{1}{\Delta} 
\big[ \lambda_1 a_1 \lambda_2 a_2 \, I_1 I_2 \, \big( I_2 K_1 - I_1 K_2 \big)
\\[2mm] \hspace{45mm} + \lambda_1 a_1 \, I_1^2 + \lambda_2 a_2 \, I_2^2 \big]
\,K_m(\kappa r) K_m(\kappa r^\prime),
& \text{if} \quad  a_1 < a_2 < (r,r^\prime), \\[3mm]
\frac{1}{\Delta} \big[ 1 + \lambda_2a_2 \,I_2K_2 \big] 
\,I_m(\kappa r) K_m(\kappa r^\prime)
- \frac{1}{\Delta} \big[ \lambda_2 a_2 \, K_2^2 \big] 
\,I_m(\kappa r) I_m(\kappa r^\prime),
& \text{if} \quad  r < a_1 < r^\prime < a_2, \\[3mm]
\frac{1}{\Delta} \,I_m(\kappa r) K_m(\kappa r^\prime),
& \text{if} \quad  r < a_1 < a_2 < r^\prime, \\[3mm]
\frac{1}{\Delta} \big[ 1 + \lambda_1a_1 \,I_1K_1 \big]
\,I_m(\kappa r) K_m(\kappa r^\prime)
- \frac{1}{\Delta} \big[ \lambda_1a_1 \,I_1^2 \big]
\,K_m(\kappa r) K_m(\kappa r^\prime),
& \text{if} \quad  a_1 < r < a_2 < r^\prime,
\end{cases}
\label{g0-sol}
\end{align}
where we have used the abbreviation
\begin{equation}
\Delta = 1 + \lambda_1 a_1 \, I_1 K_1 + \lambda_2 a_2 \, I_2 K_2 
+ \lambda_1 a_1 \lambda_2 a_2 \, I_1 K_2 \, \big( I_2 K_1 - I_1 K_2 \big).
\label{deno-delta}
\end{equation}
We have used the notation 
$I_{1,2} \equiv I_m(\kappa a_{1,2})$
and
$K_{1,2} \equiv K_m(\kappa a_{1,2})$.
In the other regions not quoted above
the solution is determined by using the reciprocal symmetry,
$g_m^{(0)}(r,r^\prime;\kappa) = g_m^{(0)}(r^\prime,r;\kappa)$,
in the Green's function.

Using the above expressions we can evaluate
\begin{equation}
g_m^{(0)}(a_1,a_2;\kappa) 
= \frac{1}{\Delta} \, I_1 K_2.
\label{g0-val}
\end{equation}
The relevant first derivatives are evaluated 
using the averaging prescription described in~\cite{Cavero:},
which is not necessary in either the Dirichlet or weak limit,
as
\begin{subequations}
\begin{align}
\bigg\{ \frac{\partial}{\partial r} g_m^{(0)}(r,r^\prime;\kappa) 
\bigg\}_{r=a_1,r^\prime=a_2} &= \frac{\kappa}{\Delta} 
\left[ I_1^\prime K_2 + \frac{\lambda_1}{2 \kappa} \,I_1 K_2 \right],
\\
\bigg\{ \frac{\partial}{\partial r} g_m^{(0)}(r,r^\prime;\kappa) 
\bigg\}_{r=a_2,r^\prime=a_1} &= \frac{\kappa}{\Delta}
\left[ I_1 K_2^\prime - \frac{\lambda_2}{2 \kappa} \,I_1 K_2 \right],
\end{align}\label{g0-der}\end{subequations}
where we have used a prime to denote the derivative of 
the modified Bessel function with respect to the argument.
The derivatives acting on the second variable in the Green's can 
be deduced using the symmetry of the Green's function.
The relevant second derivatives are evaluated to be
\begin{eqnarray}
\bigg\{ \frac{\partial}{\partial r} \frac{\partial}{\partial r^\prime} 
        g_m^{(0)}(r,r^\prime;\kappa) \bigg\}_{r=a_1,r^\prime=a_2}
&=& \frac{\kappa^2}{\Delta} \left[ I_1^\prime K_2^\prime 
+ \frac{\lambda_1}{2 \kappa} \,I_1 K_2^\prime 
- \frac{\lambda_2}{2 \kappa} \,I_1^\prime K_2
- \frac{\lambda_1}{2 \kappa} \frac{\lambda_2}{2 \kappa} \,I_1 K_2 \right].
\label{g0-2-der}
\end{eqnarray}
The above evaluations used the Wronskian,
$\left[I_m(x) K_m^\prime(x) - I_m^\prime(x) K_m(x)\right] = -1/x$,
satisfied by the modified Bessel functions.

\section{Interaction energy}

In terms of the reduced Green's function defined 
in eq.~\eqref{g0-sol} we can write 
the interaction energy, to the leading order, in eq.~\eqref{dE12} as
\begin{equation}
\frac{E_{12}^{(2)}}{L_z} = \frac{1}{(2\pi)^2} 
\sum_{m=-\infty}^{\infty} \sum_{n=-\infty}^{\infty}
(\tilde{h}_1)_{m-n} (\tilde{h}_2)_{n -m} \,L_{mn}^{(2)},
\label{DE12-hhL}
\end{equation}
where $(\tilde{h}_i)_m$ are the Fourier transforms of the functions 
$h_i(\theta)$, which describe the corrugations on the cylinders,
and are defined as
\begin{equation}
(\tilde{h}_i)_m = \int_0^{2\pi} d\theta \,e^{-im\theta} \,h_i(\theta).
\end{equation}
The matrix (or the kernel) $L_{mn}^{(2)}$ in eq.~\eqref{DE12-hhL}
is expressed in the form 
\begin{equation}
L_{mn}^{(2)} = - \frac{1}{4\pi} \int_0^\infty \kappa \,d\kappa
\,I_{mn}^{(2)}(a_1,a_2;\kappa),
\label{L2=I2-cyl}
\end{equation}
where $\kappa^2 = k_z^2 - \omega^2 = k_z^2 + \zeta^2$,
after switching to imaginary frequencies by a Euclidean
rotation, $\omega \rightarrow i \zeta$.
The related matrix $I_{mn}^{(2)}(a_1,a_2;\kappa)$ is expressed
as derivatives of the reduced Green's function in the form 
\begin{equation}
I_{mn}^{(2)}(a_1,a_2;\kappa) 
= \lambda_1 \lambda_2 
\frac{\partial}{\partial r} \,\frac{\partial}{\partial \bar{r}}
    \Big[ \,r \,\bar{r} \,g_m^{(0)}(r,\bar{r};\kappa) 
       \,g_n^{(0)}(\bar{r},r;\kappa) \Big]\bigg|_{\bar{r}=a_1, r=a_2}.
\label{I2=ddgg}
\end{equation}
The reciprocal symmetry in the Green's function leads to 
the following symmetry in the above kernel:
\begin{equation}
I_{mn}^{(2)}(a_1,a_2;\kappa)
= I_{nm}^{(2)}(a_2,a_1;\kappa).
\label{I2-sym}
\end{equation}
The expression in eq.~\eqref{I2=ddgg} is evaluated 
using eqs.~\eqref{g0-val}, \eqref{g0-der}, and \eqref{g0-2-der}:
\begin{eqnarray}
I_{mn}^{(2)}(a_1,a_2;\kappa)
&=& \frac{\lambda_1 \lambda_2}{\Delta \tilde{\Delta}}
\bigg[
I_1 K_2 \,\tilde{I}_1 \tilde{K}_2
+ \kappa a_1 \,I_1 K_2 \, \Big( \tilde{I}_1^\prime \tilde{K}_2 
  + \frac{\lambda_1}{2\kappa} \,\tilde{I}_1 \tilde{K}_2 \Big)
+ \kappa a_1 \, 
\Big( I_1^\prime K_2 + \frac{\lambda_1}{2\kappa} \,I_1 K_2 \Big)
\,\tilde{I}_1 \tilde{K}_2
\nonumber \\ && \hspace{10mm}
+ \, \kappa a_2 \,I_1 K_2 \, \Big( \tilde{I}_1 \tilde{K}_2^\prime 
  - \frac{\lambda_2}{2\kappa} \,\tilde{I}_1 \tilde{K}_2 \Big)
+ \kappa a_2 \, 
\Big( I_1 K_2^\prime - \frac{\lambda_2}{2\kappa} \,I_1 K_2 \Big)
\,\tilde{I}_1 \tilde{K}_2
\nonumber \\ && \hspace{10mm}
+ \,\kappa a_1 \kappa a_2
\Big( I_1 K_2^\prime - \frac{\lambda_2}{2\kappa} \,I_1 K_2 \Big)
\Big( \tilde{I}_1^\prime \tilde{K}_2 
      + \frac{\lambda_2}{2\kappa} \,\tilde{I}_1 \tilde{K}_2 \Big)
\nonumber \\ && \hspace{10mm}
+ \kappa a_1 \kappa a_2
\Big( I_1^\prime K_2 + \frac{\lambda_1}{2\kappa} \,I_1 K_2 \Big)
\Big( \tilde{I}_1 \tilde{K}_2^\prime
      - \frac{\lambda_2}{2\kappa} \,\tilde{I}_1 \tilde{K}_2 \Big)
\hspace{15mm}
\nonumber \\ && \hspace{10mm}
+\, \kappa a_1 \kappa a_2
\Big( I_1^\prime K_2^\prime
+ \frac{\lambda_1}{2 \kappa} \,I_1 K_2^\prime
- \frac{\lambda_2}{2 \kappa} \,I_1^\prime K_2
- \frac{\lambda_1}{2 \kappa} \frac{\lambda_2}{2 \kappa} \,I_1 K_2 \Big)
\, \tilde{I}_1 \tilde{K}_2
\nonumber \\ && \hspace{10mm}
+\, \kappa a_1 \kappa a_2 \,I_1 K_2
\Big( \tilde{I}_1^\prime \tilde{K}_2^\prime
+ \frac{\lambda_1}{2 \kappa} \,\tilde{I}_1 \tilde{K}_2^\prime
- \frac{\lambda_2}{2 \kappa} \,\tilde{I}_1^\prime \tilde{K}_2
- \frac{\lambda_1}{2 \kappa} \frac{\lambda_2}{2 \kappa} 
  \,\tilde{I}_1 \tilde{K}_2 \Big)
\bigg],
\label{I2cyl}
\end{eqnarray}
where we have used the notation in which
the modified Bessel function with a tilde on it is of order $n$
and that without a tilde is of order $m$. 
The tilde on $\Delta$ means that we use the corresponding 
modified Bessel functions in eq.~\eqref{deno-delta}.

\subsection{Dirichlet Limit}

For the case of the Dirichlet limit ($a \lambda_{1,2} \gg 1$)
the expression in eq.~\eqref{I2cyl} takes the relatively simple form
\begin{equation}
I_{mn}^{(2)D}(a_1,a_2;\kappa)
= - \frac{1}{a_1 a_2} \frac{1}{[I_2 K_1 - I_1 K_2]} 
\frac{1}{[\tilde{I}_2 \tilde{K}_1 - \tilde{I}_1 \tilde{K}_2]} 
= - \frac{1}{a_1 a_2} 
\frac{1}{D_m(\alpha;\kappa R)}
\frac{1}{D_n(\alpha;\kappa R)}
\label{I2-D}
\end{equation}
where the superscript $D$ stands for Dirichlet,
and for convenience we have introduced the function
\begin{equation}
D_m(\alpha;x) = I_m\big(x[1 + \alpha]\big) K_m\big(x[1 - \alpha]\big)
- I_m\big(x[1 - \alpha]\big) K_m\big(x[1 + \alpha]\big),
\end{equation}
and two related variables:
$R=(a_1 + a_2)/2$, which is the mean radius of the 
two cylinders under consideration, and $\alpha = a/2R$, which by definition
is less than unity. We note that
\begin{equation}
\frac{a^2}{a_1 a_2} = \frac{4 \,\alpha^2}{(1-\alpha^2)}.
\end{equation}

Two cylinders with very large radius,
such that $a_i \rightarrow \infty$ with $a$ kept fixed,
will simulate a parallel plate in the region of small 
variations in the angle ($\theta \rightarrow 0$).
This corresponds to $m,n \rightarrow \infty$,
such that $m/R$ is finite and $\alpha \rightarrow 0$.
These limits are compatible with the leading uniform asymptotic
approximants to the modified Bessel functions for large orders,
see for eg.~\cite{maias},
\begin{align}
I_m(m z) \sim \sqrt{\frac{t}{2\pi m}} \,e^{m \eta (z)}
\qquad \text{and} \qquad
K_m(m z) \sim \sqrt{\frac{\pi t}{2 m}} \,e^{-m \eta (z)},
\qquad m\to\infty,
\label{uae}
\end{align}
where
\begin{equation}
t = \frac{1}{\sqrt{1 + z^2}}
\qquad \text{and} \qquad
\eta (z) = \sqrt{1 + z^2} + \text{ln} \frac{z}{1 + \sqrt{1 + z^2}}.
\end{equation}
Using the above asymptotic behaviors and neglecting terms 
of order $\alpha$ we can deduce, for large order $m$,
\begin{equation}
\frac{4 \,\alpha^2}{(1 - \alpha^2)}
\frac{1}{D_m(\alpha;\kappa R)} \frac{1}{D_n(\alpha;\kappa R)}
\sim \frac{\kappa_m a}{\sinh \kappa_m a} \frac{\kappa_n a}{\sinh \kappa_n a},
\label{uae-dd}
\end{equation}
where we have denoted $\kappa_m^2 = \kappa^2 + (m/R)^2$
and $\kappa_n^2 = \kappa^2 + (n/R)^2$.
Using the above limiting form in eq.~\eqref{I2-D},
and after interpreting $m/R \rightarrow k$ as the Fourier transform
of the coordinate containing the corrugations on the plates,
we reproduce the expression derived for the corrugated plates 
in \cite{Cavero:}.

Using eq.~\eqref{I2-D} in eq.~\eqref{L2=I2-cyl} the 
$L$-matrix in the Dirichlet limit takes the form 
\begin{equation}
L_{mn}^{(2)D} = \frac{1}{a^2} \frac{1}{4\pi}
\int_0^\infty \kappa \, d\kappa \frac{a^2}{a_1 a_2}
\frac{1}{D_m(\alpha;\kappa R)} \frac{1}{D_n(\alpha;\kappa R)},
\label{L2D-cyl}
\end{equation}
which also leads to the corresponding result for the corrugated plates.

\subsection{Weak coupling Limit}

For the case of weak coupling limit ($a \lambda_{1,2} \ll 1$) the 
expression in eq.~\eqref{I2cyl} takes the form
\begin{equation}
I_{mn}^{(2)W}(a_1,a_2;\kappa)
= \lambda_1 \lambda_2 
\frac{\partial}{\partial a_1} a_1 \frac{\partial}{\partial a_2} a_2
\Big[ I_m(\kappa a_1) K_m(\kappa a_2) I_n(\kappa a_1) K_n(\kappa a_2) \Big],
\label{I2-W}
\end{equation}
where $W$ stands for weak coupling limit.
Using the above expression for $I$-matrix in eq.~\eqref{L2=I2-cyl}
we can write the $L$-matrix in the weak coupling limit as
\begin{equation}
L_{mn}^{(2)W} = - \frac{\lambda_1 \lambda_2}{4\pi}
\frac{\partial}{\partial a_1} a_1 \frac{\partial}{\partial a_2} 
\frac{1}{a_2} \,F_{mn}\Big(\frac{a_1}{a_2}\Big),
\label{L2-W-cyl}
\end{equation}
where we have introduced the function
\begin{equation}
F_{mn}(\beta) = \int_0^\infty x\,dx \,K_m(x) K_n(x) I_m(\beta x) I_n(\beta x).
\label{Fmn}
\end{equation}
Taking the uniform asymptotic approximants of the modified Bessel functions,
see eq.~\eqref{uae}, in the above expression, we can reproduce 
the corresponding result for the corrugated plates in \cite{Cavero:}.

It is possible to convert the above integral into a single sum 
using the technique described in \cite{Milton:2007wz}.
We replace those modified Bessel functions which are well defined
at the origin, $I_m$, with their power series expansions, then perform 
the integral using
\begin{equation}
\int_0^\infty x\,dx \, x^{m+n+2k} K_m(x)K_n(x)
= \frac{1}{2} \,\frac{2^{m +n+2k} (m+n+2k+1)!}{k! (k+m)! (k+n)! (k+m+n)!},
\quad m\geq 0,n\geq 0,k\geq 0,
\end{equation}
which leaves eq.~\eqref{Fmn} in terms of two sums.
One of the sums can be performed after regrouping the terms, and leads to 
\begin{eqnarray}
F_{mn}(\beta) &=& \frac{1}{2} \beta^{m+n} \sum_{k=0}^{\infty} \beta^{2k}
\frac{1}{(m+n+2k+1)!} \sum_{k^\prime=0}^{k} 
\frac{k!(k+m)!(k+n)!(k+m+n)!}
{k^\prime!(k^\prime+m)!(k-k^\prime)!(k-k^\prime+n)!}
\nonumber \\ 
&=& \frac{1}{2} \sum_{k=0}^{\infty} \frac{\beta^{2k+m+n}}{(2k+m+n+1)},
\qquad m\geq 0,n\geq 0.
\end{eqnarray}
Substituting the above expression into eq.~\eqref{L2-W-cyl},
and taking the derivatives with respect to $a_1$ and $a_2$, 
we immediately perform the sum, leading to
\begin{equation}
L_{mn}^{(2)W} = \frac{\lambda_1 \lambda_2}{8\pi} \frac{1}{a_2^2}
\frac{\partial}{\partial \beta} \bigg[\frac{\beta^{m+n+1}}{1 - \beta^2}\bigg]
= - \frac{\lambda_1 \lambda_2}{16\pi} \frac{1}{a^2}
\,\alpha^2 \frac{\partial}{\partial \alpha}
\bigg[ \frac{1}{\alpha} \Big(\frac{1-\alpha}{1+\alpha}\Big)^{m+n}
(1 - \alpha^2) \bigg],
\qquad m\geq 0,n\geq 0,
\label{L2W-cyl}
\end{equation}
where we have denoted $\beta = a_1/a_2$ for convenience.
Interpreting $m/R \rightarrow k_1$ and $n/R \rightarrow k_2$,
and taking the limit $m\rightarrow\infty$, $n\rightarrow\infty$,
while keeping $k_{1,2}$ fixed, we obtain the expression for the 
$L$-kernel for corrugated plates in the weak limit~\cite{Cavero:}.

\section{Sinusoidal corrugations}

\begin{figure}
\begin{center}
\includegraphics[width=50mm]{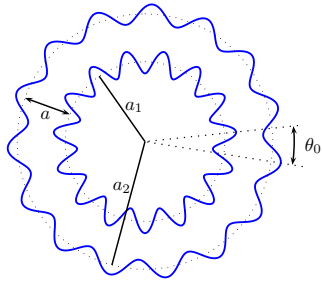}
\caption{Non-contact gears: Concentric corrugated cylinders 
with the same corrugation frequency, $\nu=15$, on each cylinder. 
$\theta_0$ is the angular shift between the gears. }
\label{corru-cyl}
\end{center}
\end{figure}

For the particular case of sinusoidal corrugations,
as described in figure \ref{corru-cyl}, we will have
\begin{subequations}
\begin{eqnarray}
h_1(\theta) &=& h_1 \sin [\nu(\theta + \theta_0)],
\\
h_2(\theta) &=& h_2 \sin [\nu \theta],
\end{eqnarray}\label{sin-cyl}\end{subequations}
where $h_{1,2}$ are the corrugation amplitudes and
$\nu$ is the frequency associated with the 
corrugations. Necessarily, $\nu$ must be a positive integer.
The Fourier transforms, $(\tilde{h}_i)_m$,
corresponding to the above corrugations are
\begin{equation}
(\tilde{h}_1)_m = h_1 \frac{2\pi}{2i}
\Big[ \,e^{i\nu\theta_0} \delta_{m,\nu}
- e^{-i\nu\theta_0} \delta_{m,-\nu} \Big].
\label{h-four}
\end{equation}

In general the corrugation frequencies of the two cylinders
can be different. However, we note that,
to the leading order, the interaction energy
gets contributions only when both cylinders have the same frequency.

Using the above expression in eq.~\eqref{DE12-hhL}
and using the symmetry property of $I^{(2)}_{mn}$ in eq.~\eqref{I2-sym}, 
which further lets us deduce $L^{(2)}_{mn} =L^{(2)}_{nm}$,
we can write
\begin{equation}
\frac{E_{12}^{(2)}}{L_z} 
= \cos(\nu\theta_0) \,\frac{h_1 h_2}{2} 
\sum_{m=-\infty}^{\infty} L^{(2)}_{m,m+\nu}
= - \cos(\nu\theta_0) \,\frac{h_1 h_2}{8\pi}
\sum_{m=-\infty}^{\infty} \int_0^\infty \kappa \,d\kappa
\, I^{(2)}_{m,m+\nu}(a_1,a_2;\kappa)
\label{DE12-L}
\end{equation}
where the kernel has been explicitly evaluated in eq.~\eqref{I2cyl}.

\subsection{Dirichlet Limit}

In the Dirichlet limit the interaction energy in eq.~\eqref{DE12-L}
can be expressed in the form
\begin{equation}
\frac{E_{12}^{(2)}}{2\pi R\,L_z}
= \cos(\nu\theta_0) \,\frac{\pi^2}{240 \,a^3} 
\,\frac{h_1}{a} \frac{h_2}{a} \, B_\nu^{(2)D}(\alpha)
\label{DE12-D-cyl}
\end{equation}
where we have divided by a factor of $2\pi R$, which is the 
mean circumference in the direction of corrugations. We have also defined 
a suitable function $B_\nu^{(2)}(\alpha)$ to make it convenient 
to compare our results with those obtained in the PFA limit and
with those for corrugated plates in the appropriate limits.
We define
\begin{equation}
B_\nu^{(2)D} (\alpha)
= \frac{15}{\pi^4} \sum_{m=-\infty}^{\infty} 8 \alpha^3 \int_0^\infty x \,dx
\frac{4 \, \alpha^2}{(1 - \alpha^2)}
\frac{1}{D_m(\alpha;x)}
\frac{1}{D_{m+\nu}(\alpha;x)}.
\end{equation}
\begin{figure}
\begin{center}
\includegraphics[width=80mm]{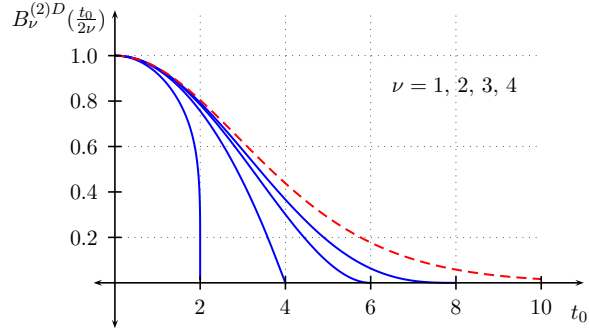}
\caption{
Dirichlet limit: Plots of $B_\nu^{(2)D}(\frac{t_0}{2\nu})$ versus $t_0$,
for $t_0 < 2\nu$ and fixed $\nu$.
The dashed curve is the corresponding plot for corrugated plates
which is approached by the corrugated cylinders for larger values
of $\nu$. }
\label{diri-fig}
\end{center}
\end{figure}
Using eq.~\eqref{uae-dd} it is straightforward to verify the 
corrugated plate limit of the above expression.
The function $B_\nu^{(2)D} (\alpha)$ has been plotted with respect
to $t_0 = 2\alpha\nu$ in figure \ref{diri-fig}. The redefined
parameter helps us compare our results with the corrugated plates.
We note that for larger values of $\nu$ the plots approach
the curve for the corrugated plates very quickly.
We note that $B_\nu^{(2)D}(1)=0$, because for $\alpha =1$
the radius of the inner cylinder approaches zero. Thus, it is pointless
to consider the regime $\alpha >1$.
We also note that $B_\nu^{(2)D}(0)=1$, which then implies the PFA limit.

The Casimir torque per unit area, for the Dirichlet case, can thus 
be evaluated, using eq.~\eqref{DE12-D-cyl} in eq.~\eqref{torE}, to be
\begin{equation}
\frac{{\cal T}^{(2)D}}{2\pi R\,L_z}
= \nu \sin(\nu\theta_0) \,\frac{\pi^2}{240 \,a^3} 
\,\frac{h_1}{a} \frac{h_2}{a} \, B_\nu^{(2)D}(\alpha).
\end{equation}

\subsection{Weak coupling limit}

The evaluation for the $L$-matrix in the weak limit in eq.~\eqref{L2W-cyl}
is valid for positive indices only. 
Therefore, we begin by rewriting the expression for the interaction energy
in eq.~\eqref{DE12-L} in the form 
\begin{equation}
\frac{E_{12}^{(2)W}}{L_z} = \cos(\nu\theta_0) \,\frac{h_1 h_2}{2}
\bigg[ \,2 \sum_{m=0}^{\infty} L^{(2)W}_{m,m+\nu} 
+ \sum_{m=1}^{\nu -1} L^{(2)W}_{m,\nu -m} \bigg],
\end{equation}
where the finite sum is interpreted as zero when $\nu=1$.
We have used the symmetry property mentioned before 
eq.~\eqref{DE12-L}, and further used 
$L^{(2)W}_{m,n}=L^{(2)W}_{-m,n}=L^{(2)W}_{m,-n}=L^{(2)W}_{-m,-n}$,
which can be deduced from eq~\eqref{I2-W} using the Bessel function
property $I_{-m}(x)=I_{m}(x)$ and $K_{-m}(x)=K_{m}(x)$.
After substituting the $L$-matrix, derived in eq.~\eqref{L2W-cyl},
into the above equation we can immediately perform the sums to yield
\begin{equation}
\frac{E_{12}^{(2)W}}{2\pi R\,L_z}
= \cos(\nu\theta_0) \, \frac{\lambda_1 \lambda_2}{32 \pi^2\,a}
\,\frac{h_1}{a} \frac{h_2}{a} \, B_\nu^{(2)W}(\alpha),
\label{DE12-W-cyl}
\end{equation}
where we have defined the function
\begin{equation}
B_\nu^{(2)W}(\alpha)
= - \frac{\alpha^3}{2} \frac{\partial}{\partial \alpha}
\bigg[ \frac{1}{\alpha^2} \left( \frac{1-\alpha}{1+\alpha} \right)^\nu
(1-\alpha^2) (1 + 2\alpha\nu + \alpha^2) \bigg].
\label{bnu-W}
\end{equation}
The Casimir torque per unit area, for the weak coupling limit, can thus
be evaluated, using eq.~\eqref{DE12-W-cyl} in eq.~\eqref{torE}, to be
\begin{equation}
\frac{{\cal T}^{(2)W}}{2\pi R\,L_z}
= \nu \sin(\nu\theta_0) \, \frac{\lambda_1 \lambda_2}{32 \pi^2\,a}
\,\frac{h_1}{a} \frac{h_2}{a} \, B_\nu^{(2)W}(\alpha).
\end{equation}

We note that $B_\nu^{(2)W}(0) =1$. This verifies that
the above result satisfies the proximity force theorem.
As in the Dirichlet case,
we note that $B_\nu^{(2)W}(1)=0$, because for $\alpha =1$
the radius of the inner cylinder approaches zero.
The above result should also yield the result for 
corrugated parallel plates in
the limit $\nu \rightarrow \infty$, $a_{1,2} \rightarrow \infty$,
$R \rightarrow \infty$, such that $a$ and $\nu/R$ is finite.
Also, in this limit $\alpha \rightarrow 0$.
Recalling the corrugated plates parameter~\cite{Cavero:},
$k_0a \rightarrow \nu a/R = 2\nu\alpha \equiv t_0$,
we note that the limit to corrugated plates is achieved by 
taking the limit $\nu\rightarrow\infty$ with $t_0$ kept fixed.
To this end we rewrite eq.~\eqref{bnu-W} in terms of $t_0$ as 
\begin{equation}
B_\nu^{(2)W}\Big(\frac{t_0}{2\nu}\Big)
= - \frac{t_0^3}{2} \frac{\partial}{\partial t_0}
\bigg[ \frac{1}{t_0^2} \left( 1 - \frac{t_0}{2\nu} \right)^\nu
\left( 1 + \frac{t_0}{2\nu} \right)^{-\nu}
\left(1-\frac{t_0^2}{4\nu^2} \right)
\left(1 + t_0 + \frac{t_0^2}{4\nu^2} \right) \bigg]
\end{equation}
\begin{figure}
\begin{center}
\includegraphics[width=80mm]{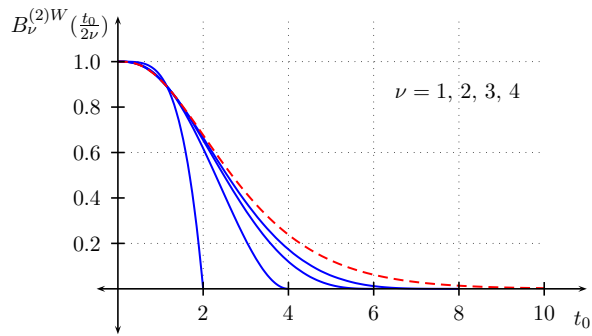}
\caption{
Weak coupling limit: Plots of $B_\nu^{(2)W}(\frac{t_0}{2\nu})$ versus $t_0$,
for $t_0 < 2\nu$ and fixed $\nu$.
The dashed curve is the corresponding plot for corrugated plates
which is approached by the corrugated cylinders for larger values
of $\nu$. }
\label{weak-fig}
\end{center}
\end{figure}
in which the $\nu\rightarrow\infty$ limit can be immediately taken to yield
\begin{equation}
\lim_{\nu\rightarrow\infty} B_\nu^{(2)W}\Big(\frac{t_0}{2\nu}\Big)
= - \frac{t_0^3}{2} \frac{\partial}{\partial t_0}
\bigg[ \frac{1}{t_0^2} (1 + t_0) \,e^{-t_0} \bigg]
= \frac{t_0^3}{2} \,\frac{\partial^2}{\partial t_0^2}
\bigg[ \frac{1}{t_0} \,e^{-t_0} \bigg],
\end{equation}
which matches the result for the corrugated plates exactly.
The function $B_\nu^{(2)W} (\alpha)$ has been plotted with respect
to $t_0$ in figure \ref{weak-fig}.
As in the case of Dirichlet case, we note that for larger values of 
$\nu$ the plots approach the plot for corrugated plates very quickly.

\section{Conclusion and future directions}

We have evaluated the Casimir torque between two concentric
corrugated cylinders perturbatively in the corrugation amplitude
for the scalar case. 
We have derived explicit expressions for the case when the 
cylinders have sinusoidal corrugations. 
Nonzero contributions in the leading order requires the 
corrugation frequencies on the two cylinders to be identical. 
Our results for the Casimir torque reproduce
the results for the lateral force on corrugated parallel plates
in the limit of large radii and small corrugation wavelengths.

Extension of the calculation to the next-to-leading order
in the spirit of~\cite{Cavero:} is in progress.  
We recall that it was possible to evaluate the lateral force between 
corrugated parallel plates, in the weak limit, exactly in terms
of a single integral~\cite{Cavero:}. The corresponding exact 
result for the Casimir torque between corrugated
cylinders, in the weak limit, is being sought.
Generalization to the electromagnetic case is most important
and we intend to attempt it next.
Generalization of the calculation to the case of a rack and pinion
arrangement is another possible extension.
We hope that these calculations will have application in the design
of practical nanomechanical devices.

\begin{acknowledgments}
We thank Jef Wagner for extensive collaborative assistance
throughout this project.
We thank the US National Science Foundation (Grant No. PHY-0554926)
and the US Department of Energy (Grant No. DE-FG02-04ER41305)
for partially funding this research.
ICP would like to thank the French National Research Agency (ANR)
for support through Carnot funding.
\end{acknowledgments}

\end{document}